\documentstyle[12pt, emlines]{article}

\newcommand{\f}{\varphi}

\newcommand{\mb}{\bf}
\newcommand{\w}{\omega}

\newcommand{\be}{\begin{equation}}
\newcommand{\ee}{\end{equation}}

\title{ Renormalization Group Approach to the
Problem of Flow Through Irregular Packed Beds}
  \author{Dmitri  Volchenkov and Ricardo Lima\\
  \small
   \em  CNRS, Centre de Physique Theorique, 13288 Marseille
  Cedex 09, Luminy Case 907, France \\
\small
   \em volchen@cptsg2.univ-mrs.fr, lima@cptsg2.univ-mrs.fr
}

\begin{document}
\thispagestyle{empty}

\noindent
\maketitle

\begin{abstract}

We have calculated the asymptotics of
Green's function of the differential equation of nonlinear diffusion
in the microscopic range with strong porosity fluctuations
in the problem of flow through irregular packed beds for
the arbitrary dimension of space and arbitrary
porosity fluctuations covariance.

 \end{abstract}


\section{Introduction}

Properties of flows in porous media are of great interest of modern industry and technology.
In spite of practical importance and long-standing attention paid to this
problem there is still no solidarity on  model of the flow through the packed beds.

One can point out an approach of a single particle diffusing with random dynamics in a
 Poissel velocity field, which goes back to the classical work of Taylor \cite{1}.
 Nevertheless, the experimental  measurements are still described by the usual
 diffusion equation where a modified dispersion coefficient is used;
 see \cite{2} and references therein. The problem of these considerations is of
 the essentially gaussian solutions of the diffusion equation.  The numerous
 experimental observations, \cite{2} demonstrates clearly that the previous approach
 needs to be supplemented.

 One of the first attempts to overcome this limitation is proposed in \cite{3}, where
 a Markov random process is used to model the diffusion of impurities
 in a turbulent stream. In \cite{4} the stochastic field was introduced to model the
 porosity instead of the diffusion process itself.
Their result for the  one-dimensional Fick law of  diffusion
 for some marker transported by a carrier fluid through an irregularly but statistically
 homogeneous packed
 column leads to the appearance in the equation of the mean concentration of
 a time diffusion term. Together with a source term they are to be responsible for
 the  departure of the solutions from gaussian.

 In  this paper we keep on study  the diffusion equation of the Fick-type in which the
 porosity $\varepsilon$ fluctuates strongly along  the column axis.
There are many methods of defining porosity; see \cite{5} and references therein. Each of
them replaces the complex network of voids with a single number
that represents an average property. In the present work we use porosity as a point
quantity in a porous medium and define an elementary  volume to characterize a medium macroscopically following
\cite{6} and \cite{7}. The size of the elementary volume around a point $P$ we suppose to be smaller than
the total medium so that  it can represent a fluid flow at $P$. We suppose  also that there is enough pores to
allow statistical averaging. Since porosity varies, the maximum length is the characteristic length $\xi$
that indicates the rate of change of porosity, and the minimum length $w$ is the pore size.

The porosity is directly related to the size of the pores relative to the matrix. Considering the problem
 of packed column, we denote
 by $x$ the axial coordinate of the column, $0\leq x\leq L,$ then
 the fraction of the column that is voids defines the averaging value of $\varepsilon$,
 $0\leq \varepsilon\leq 1.$ If $P$ be a point inside a porous medium on the column axis, see Fig.1.,
 surrounded by a layer of volume $V(l_i)$, one can define a ratio
 $$\varepsilon_i=\frac {V_v(l_i)}{V(l_i)}$$
 in which $V_v(l_i)$ is the volume of void space within volume $V(l_i)$.

 Consider a sequence of values of $l_i$ such that $l_1>l_2>l_3>\ldots .$
 For large values of $V(l_i)$ the ratio $\varepsilon_i$ may change gradually as $V(l_i)$
 gets smaller. As $l$ falls below a certain value $\xi$ there will be large fluctuations
 in porosity when $V(l)$ is approaching the dimension of the pores (this $V(\xi)$ may
 be taken as the elementary volume). Below this value the average value of porosity
 has no meaning, then it is natural to consider the porosity as a function of $x$,
 where $x$ be a distance apart from the point $P.$

 Flow phenomena  for $x>\xi$ is well investigated in the porous medium theory;
 see for example \cite{5}  for a review.  Below $\xi$,  for $w\leq x < \xi$ flow properties are
 subject to strong microscopic effects and usually left beyond the investigations due to an extremely large
 fluctuations in porosity when $x$ is approaching the dimension of the pores. However,
 namely this interval of scales is of
 great interest for the variety of problems in
 chemical engineering and environmental studies.

 We continue to study the microscopic properties of flows through a porous media
 with strongly fluctuative porosity using the quantum field theory renormalization group method (RG)
 and $4-\epsilon$ expansion \cite{8}. The technique which is developed in what following is similar to one of \cite{9}
devoted to waves propagation in a randomly  inhomogeneous medium with    strongly
developed fluctuations.

 In Sec. 2,  the problem is stated and formulated in the quantum-field formalism.  In Sec. 3,
 the Green's function for the equation of nonlinear diffusion is constructed for
 the case of  gaussian distributed fluctuative porosity. We have introduced the effective action functional
 which  is equivalent to the problem discussed under the certain conditions in Subsec. 3.1.
 Investigating the properties of the theory, we have justified the result of \cite{4} on the
 inclusion of a "time-diffusive" term  into the equation of nonlinear diffusion in case of fluctuating porosity
 for  arbitrary dimension of space and for arbitrary stochastic process originating the porosity fluctuations.

 In Subsec. 3.3 and Subsec. 3.4, the procedure of ultraviolet renormalization of the theory with
 the "time-diffusive" term included  is considered and critical dimensions including anomalies are
 calculated. The  results obtained for static correlation functions meet the well-known  empirical laws
 of Richardson and Kolmogorov. It is shown  that for  the parameters taking values within the physical
 domain (positive viscosity, {\em etc.}), the renormalization-group equations have  a manifold
 of  infrared (IR) attractive fixed points. This implies the presence of a universal scaling regime in
 the IR domain of the microscopical region of fluctuative porosity for some range of values of
 physical parameters.

 The main result of the present paper is the detailed description of  the effect of
 strongly fluctuating porosity on the form of  static ( {\em i.e.,} of equal time)
 as well as of  dynamic Green's functions  of the differential equation of nonlinear diffusion.

 Similar to the   problems of nonlinear diffusion and of  chemically active scalar
 admixture, the problem  of the flow  through the irregular packed beds  with strongly fluctuative porosity,
in the general case, is of  an infinite number of coupling constants, {\em i.e.,}
no one of the statistical momenta of higher order can be omitted from the consideration.
 However,  in contrast with the problems  mentioned  any  of these terms does not contain
  UV-divergences, so that the  UV renormalization procedure   as well as  the critical behavior of
the quantities  are unsensitive to certain  hypotheses on the porosity function  as well as
the certain statistical distribution for porosity in the microscopical range.

 It turns out to be that fluctuative porosity does not affect
 on the universal properties of the static Green's functions, {\em i.e.,}
 it does not change  the indices  of empirical laws of Richardson and Kolmogorov.
 But  porosity changes the amplitude factors of  static correlation, {\em i.e.,}
  the Kolmogorov constant and the effective diffusivity coefficient.

Fluctuative porosity does affect the dynamic Green's functions
providing the damping retarded time-spectrum.

 Qualitatively, for the concentration one has an outgoing  damping traveling wave  type solution
 which  decay rate  as well as an amplitude depend on porosity.

\section{ Formulation of the Problem}

We  shall consider the diffusion process of mixture
in a randomly inhomogeneous packed beds with strongly
fluctuative  porosity in the microscopic range $w\leq |\mb x| \ll \xi$
for the arbitrary value of space dimension $d.$

  The problem is posed as follows: To find the retarded Green's
  function (propagator), averaged over the imposed statistics
  of fluctuative porosity $\varepsilon (\mb x)$ and flow velocity
  $\mb v (\mb x, t),$ of the inhomogeneous diffusion
  equation for the concentration $u(\mb x,t )$ of some marker
  in the available fraction of the volume:
\be
 \varepsilon (\mb x)\partial_t  u({\mb x},t)+(v\partial) u({\mb x},t)+ \nu \Delta u({\mb x},t)=0
\label{eq}
\ee
in which
$\nu$ is the Fick's diffusion coefficient,
$\Delta$ is the Laplace operator,
$\mb v({\mb x},t) $ is the velocity of fluid flow.
The porosity $\varepsilon(\mb x)=\varepsilon_0 + \tilde\varepsilon ({\mb x} )$ is the sum
of constant  and random components. From the physical point of view
the value of porosity belongs to the interval $\varepsilon\in \left[0,1\right],$
so that, generally speaking,  $\tilde\varepsilon$ cannot be taken as a gaussian distributed function.
To model porosity one can use any function which values
 lie within the interval  $ \left[0,1\right];$
for a qualitative determinacy in what following we model it
by the function
\be
 \varepsilon (\mb x)= (1+ \sin c^{-1}\f( \mb x))/2,
\label{poros}
\ee
where  $c\equiv \nu/\xi$ is a parameter  with the dimensionality of velocity which characterizes
the velocity of the diffusion process in the scale  $\xi$, and
$\f( \mb x)$ is a gaussian distributed  phase function (of the same dimensionality) with zero mean value and
static correlation function of the  Ornstein-Zernike type, \cite{M}
\be
D_\f \equiv\langle\f(\mb x )\f(\mb y )\rangle=
  \eta_0\frac{\nu_0^3}{c_0^2}\frac 1{(2\pi)^d}\int d{\mb k}
 (k^2+\mu^2)^{(2-d -2 \delta)/ 2} \exp i{\mb k}({\mb x-\mb y}),
 \label{korrfi}
 \ee
 where $\mu\simeq 1/\xi$ is a mass parameter. We consider all parameters in
 (\ref{korrfi}) as bare (nonrenormalized ) parameters and supply them with the index $"0".$
  For  future convenience in (\ref{korrfi}) we have introduced
the amplitude factor by such a way that   it  meets
 the proper dimensionality of the correlation.
The  coupling constant  $\eta_0$  plays the role of  an expansion parameter in the perturbation theory
such that  the  relation $\eta_0\nu_0\equiv \Lambda^{2\delta}$ defines the characteristic  maximum
momentum scale in the theory, $\Lambda\simeq 1/w.$

The parameter of the regular expansion of   correlation functions  $2\delta $
is  the deviation of the space dimensionality from $2$.
We  demonstrate below  that under  the certain physical assumptions
the real value of  parameter $\delta $ is $\delta_r=1$.

The dragging velocity $v$  along with the  fluctuating component of
porosity   $\tilde\varepsilon$ prescribe  a  certain dimensionless parameter
in the problem which is  analogous to  the Reynolds number, $Re_{\varepsilon}=\tilde{\varepsilon} vw/\nu,$
where $\tilde{\varepsilon}$ is   characteristic porosity fluctuations in the scale
of   $w$, and $v$ is a characteristic fluctuative flow velocity in a pore.
A statistical steady state is expected
when $Re_{\varepsilon}\gg 1.$ In terms of the energy dissipation
of fluid flow  the parameter $Re_{\varepsilon}$ defines a dissipation length $l_d$ by the relation
$Re_{\varepsilon}^{4/3}\sim l_dw^{-1}. $  Starting from some values $Re_{\varepsilon}$,
the  dissipation length   falls into the range of microscopic effects, $w\ll r\ll \xi.$
It leads to the  formation of   inertial range of scales, $l_d\ll \xi,$ which is
substantially analogous to the Kolmogorov's dissipation free  range \cite{Monin}.

The basic results of the phenomenological Kolmogorov-Obukhov
 theory  \cite{11} can be generalized directly to the   problem in question: Assume that in the
 inertial range of wave numbers $(1/\xi \ll k\ll 1/ l_d)$ correlation
 functions of $\f$ depend  not only on the  mean energy  dissipation rate $W$
 of   fluid flow but also on an additional argument, the velocity
 of diffusion process, $c$. Then, assuming that the correlation functions do not depend on viscosity,
 for an equal-time pair correlation function of $\f$, one has
 \be
 D(k)\sim W^{2/3}k^{-11/3}{\cal F}(W^{2/3}k^{-2/3}c^{-2}),
 \label{hyp}
 \ee
 where ${\cal F}$ is a scaling function of dimensionless arguments.
 By the way, (\ref{hyp}) predicts the Kolmogorov's scaling ($\Delta_{\f}=-1/3$) for the stochastic
 phase field $\f$.
 This phenomenological result is taken into account in the model correlation
 (\ref{korrfi}) by choosing  $\delta_r=1$ as a real value for $\delta$. In accordance with (\ref{hyp}) and
 \cite{16}, one obtains  the scaling dimensions for  quantities in
 the amplitude factor of (\ref{korrfi}) $\Delta_c=-1/3$ and $\Delta_{\nu}=-2/3$;
for $\delta=1$ (\ref{korrfi}) provides exactly the Kolmogorovian scaling dimension for
 the correlation $\langle\f\f\rangle.$

Similarly, to describe the effect of eddy diffusion which obviously plays an essential role in
the microscopic diffusion process
  we consider the dragging velocity to be a $d$-dimensional gaussian
 vector field ${\mb v}({\mb x},t)$ of zero mean value and of covariance:
 \be
D_v\equiv \langle v_i({\mb x},t)v_j({\mb y},t')\rangle =  g_0\nu_0\frac {\delta(t'-t)}{(2\pi)^d}\int d{\mb k}
 P_{ij}(  k ) (k^2+m^2)^{-\frac d2-\frac \epsilon 2} \exp i{\mb k}({\mb x-\mb y}),
 \label{4}
 \ee
where $P_{ij}(  k )=\delta_{ij}-k_ik_j/k^2$
is the transverse projector (the fluid flow is considered as incompressible),
$g_0$ is a relevant coupling constant ( another expansion parameter in the perturbation theory),
  $1/m$ is a turbulent integral scale $\Lambda'$ which can be chosen as $\Lambda'=\Lambda$; $0<\epsilon< 2$
is another parameter of   regular expansion of Green's functions which real  ("Kolmogorov's") value
 is $\epsilon_r=2/3$.
 Note, that the velocity correlation (\ref{4}) as specified above is inherent for numerous simplified
models of turbulence introduced by Obukhov, \cite{10} and Kraichnan, \cite{11}, see
\cite{12} and references therein.

   The object of our interest is $G\equiv \langle \langle G (\f, v)\rangle \rangle,$
where $G (\f, v)$ is the Green's function of the differential equation (\ref{eq})
for $\f$ and $\mb v$ fixed  and the doubled angular brackets denote the operations
of functional averaging over the known statistics of  $\f$ and $\mb v$.

The stochastic phase factor $ \sin c^{-1}\f (\mb x)$ can be expanded in powers of  $\f$,
$$\sin c^{-1} \f (\mb x)=\sum_{n=1}^{\infty}
 \frac {(c^{-1}\f)^{2n-1}(\mb x)}{2n-1!}$$
 as well as the averaged product of two phase factors is expressed by a series of
various statistical momenta,
$$\langle \sin c^{-1}\f (\mb x)\sin c^{-1}\f (\mb y)\rangle=\sum_{n,m=1}^{\infty}
\frac{\langle(c^{-1}\f)^{2n-1} (\mb x)(c^{-1}\f)^{2m-1} (\mb y)\rangle}{2n-1!2m-1!}.$$
 In order to sidestep consideration of the onset of the
problem we compute first the Green's function of the simplified equation
with no terms proportional to $\f^k(\mb x), k>1.$ Contributions of the higher statistical momenta
then are to be  determined by various composite operators $F_k=\langle c^{-k}\f^k( \mb x)\rangle $
 which we study   in  the forthcoming sections.

We conclude this section by an explanation of the physical meaning of   propagator
$G$. The Fourier transformed propagator $G(k), k =|{\mb k}|$ is to be computed in the framework of diagrammatic
technique of Feynman. Assuming that initially the marker was concentrated in the source of frame,
$J({\mb x},t)=\delta({\mb x})f(t),$ one obtains the relevant concentration profile of the traveling wave type,
$u({\mb r},t)=u(r)f(t-r\varepsilon/v)$ by the expression
\be
u(r)=(2\pi)^{-d}\int d{\mb x} {\ }G(  k ) \exp i{\mb k \mb x}.
\label{5}
\ee
Another practically important problem is to define a concentration distribution profile
in a porous half-space. The flat boundary of the medium plays the role of a source term in the right
hand side of the equation (\ref{eq}). In case of the porous half-space
$x_1\equiv r\geq 0,$ the concentration profile $\langle u(r,t)\rangle$ inside the porous medium
is determined  by the equation (\ref{eq}) with a source term of the form $J({\mb x},t)=\delta(x_1)f(t)$
and yielded by the same formula (\ref{5}) with $d=1$.

Notice that (\ref{5}) allows one to bring about the angular integration   for arbitrary value of space
dimension $d,$
\be
u(r)=(2\pi)^{-d/2}r^{1-d/2}\int^\infty_0 dk {\ } G(k) {\ }k^{d/2} J_{d/2-1}(kr).
\label{6}
\ee
where $J_\nu$ is the Bessel function. Qualitatively, the solution  of (\ref{6})
is always an outgoing damping traveling wave of concentration  which  decay rate is of essentially
interest for practical applications. We shall discuss  the calculation of this
rate below in Section 4.

\section{Solutions for the Case of Gaussian Distributed $\tilde\varepsilon$}

To illustrate our approach we start  our consideration  with  unphysical
but the simplest case of gaussian distributed  fluctuative component
of   porosity  $\tilde\varepsilon$.  In the forthcoming sections we
recover  the results for the general case.

 \subsection{  Diagram Technique and Effective Action Functional}

In  the  present  section  we develop the  diagram technique relevant to the
problem discussed. The bare  propagator  $L_0^{-1}$ (the Green's function of the differential equation
(\ref{eq})  with no nonlinear terms proportional to $\f$ and $\mb v$) is a retarded function,
\be
\langle \psi({\mb k},t)\psi^\dagger({ - \mb k},t')\rangle_0 = L_0^{-1}(k,t-t')=\theta(t-t') e^{-\nu_0k^2(t-t')},
\label{7}
\ee
and   can be naturally pictured   as an oriented line in   diagrams. Diagrammatic
expression for the propagator  $G(\f,\mb v)$ is

\unitlength=1mm
\special{em:linewidth 0.4pt}
\linethickness{0.4pt}
\begin{picture}(146.00,43.00)
\put(14.00,27.00){\makebox(0,0)[cc]{$G(\varphi,\bf v)=$}}
\emline{31.00}{27.00}{1}{55.00}{27.00}{2}
\emline{45.00}{28.00}{3}{48.00}{27.00}{4}
\emline{48.00}{27.00}{5}{45.00}{26.00}{6}
\put(61.00,27.00){\makebox(0,0)[cc]{$+$}}
\emline{66.00}{27.00}{7}{97.00}{27.00}{8}
\emline{81.00}{27.00}{9}{81.00}{30.00}{10}
\emline{81.00}{33.00}{11}{81.00}{36.00}{12}
\emline{81.00}{39.00}{13}{81.00}{43.00}{14}
\emline{73.00}{28.00}{15}{77.00}{27.00}{16}
\emline{77.00}{27.00}{17}{73.00}{26.00}{18}
\emline{89.00}{28.00}{19}{93.00}{27.00}{20}
\emline{93.00}{27.00}{21}{89.00}{26.00}{22}
\put(107.00,27.00){\makebox(0,0)[cc]{$+$}}
\emline{115.00}{27.00}{23}{142.00}{27.00}{24}
\emline{128.00}{27.00}{25}{134.00}{29.00}{26}
\emline{134.00}{29.00}{27}{128.00}{31.00}{28}
\emline{128.00}{31.00}{29}{134.00}{34.00}{30}
\emline{134.00}{34.00}{31}{128.00}{36.00}{32}
\emline{128.00}{36.00}{33}{134.00}{39.00}{34}
\emline{134.00}{39.00}{35}{128.00}{41.00}{36}
\emline{128.00}{41.00}{37}{134.00}{43.00}{38}
\emline{137.00}{28.00}{39}{140.00}{27.00}{40}
\emline{140.00}{27.00}{41}{137.00}{26.00}{42}
\emline{121.00}{28.00}{43}{125.00}{27.00}{44}
\emline{125.00}{27.00}{45}{121.00}{26.00}{46}
\put(147.00,27.00){\makebox(0,0)[cc]{$+\ldots$}}
\end{picture}

in which the dashed tail corresponds to the field ${\mb v}$ and
 the saw-type tail
is associated with the field $\f.$ In what following functional
 averaging over $\f$ and $\mb v$
the relevant tails are connecting to each other, producing the dashed
 lines for the
velocity correlation function (\ref{4}) and the saw-type lines for the
 stochastic phase correlation function (\ref{korrfi}). Up to one-loop
order the diagram series for the averaged propagator $G$ is
forthcoming. The diagrams  are similar
to the diagrams
  of ordinary quantum field theory in which some
"physical" complex valued field $\psi^\dagger({\mb x},t)$,
$\psi({\mb x},t)$ supplied by the  bare propagator
$L_0^{-1}$ interacts with a couple of stochastic fields
 $\f ({\mb x})$ and $\mb v(\mb x,t).$

\newpage

\be
\unitlength=1mm
\special{em:linewidth 0.4pt}
\linethickness{0.4pt}
\begin{picture}(145.00,30.00)
\put(6.00,21.00){\makebox(0,0)[cc]{$G(\varphi,\bf v)=$}}
\emline{26.00}{21.00}{1}{48.00}{21.00}{2}
\emline{37.00}{22.00}{3}{41.00}{21.00}{4}
\emline{41.00}{21.00}{5}{37.00}{20.00}{6}
\put(54.00,21.00){\makebox(0,0)[cc]{$+$}}
\emline{62.00}{21.00}{7}{92.00}{21.00}{8}
\emline{67.00}{21.00}{9}{69.00}{24.00}{10}
\emline{70.00}{25.00}{11}{74.00}{27.00}{12}
\emline{75.00}{27.00}{13}{79.00}{28.00}{14}
\emline{80.00}{28.00}{15}{85.00}{27.00}{16}
\emline{87.00}{26.00}{17}{90.00}{23.00}{18}
\emline{90.00}{22.00}{19}{91.00}{21.00}{20}
\emline{92.00}{21.00}{21}{97.00}{21.00}{22}
\emline{64.00}{22.00}{23}{66.00}{21.00}{24}
\emline{66.00}{21.00}{25}{64.00}{20.00}{26}
\emline{77.00}{22.00}{27}{81.00}{21.00}{28}
\emline{81.00}{21.00}{29}{77.00}{20.00}{30}
\emline{93.00}{22.00}{31}{96.00}{21.00}{32}
\emline{96.00}{21.00}{33}{93.00}{20.00}{34}
\put(104.00,21.00){\makebox(0,0)[cc]{$+$}}
\emline{109.00}{21.00}{35}{143.00}{21.00}{36}
\emline{117.00}{21.00}{37}{120.00}{23.00}{38}
\emline{120.00}{23.00}{39}{118.00}{26.00}{40}
\emline{118.00}{26.00}{41}{123.00}{26.00}{42}
\emline{123.00}{26.00}{43}{125.00}{30.00}{44}
\emline{125.00}{30.00}{45}{127.00}{26.00}{46}
\emline{127.00}{26.00}{47}{131.00}{26.00}{48}
\emline{131.00}{26.00}{49}{129.00}{23.00}{50}
\emline{129.00}{23.00}{51}{132.00}{21.00}{52}
\emline{111.00}{22.00}{53}{113.00}{21.00}{54}
\emline{113.00}{21.00}{55}{111.00}{20.00}{56}
\emline{123.00}{22.00}{57}{126.00}{21.00}{58}
\emline{126.00}{21.00}{59}{123.00}{20.00}{60}
\emline{136.00}{22.00}{61}{139.00}{21.00}{62}
\emline{139.00}{21.00}{63}{136.00}{20.00}{64}
\put(145.00,21.00){\makebox(0,0)[cc]{$+\ldots$}}
\end{picture}
\label{8}
\ee
  The relevant action functional has the form
\be
S(\f,\mb v,\psi)= \int d{\mb x}dt\left[-\frac 12 \f D^{-1}_\f \f-\frac 12 \mb vD^{-1}_v\mb v+
\psi^\dagger\left( L_0\psi+ \f\partial_t\psi + (v\partial)\psi \right) \right] + C.C.
\label{9}
\ee
where $C.C.$ denotes a complex conjugate part, $D_\f$ and  $D_v$ are the correlation functions (\ref{korrfi}) and (\ref{4})
consequently. Averaging the
product $\psi({\mb x},t)\psi^\dagger({\mb y},t)$ over the complete set of fields
$\{\f,\mb v,\psi\}$ for the statistical weight $\exp S(\f,\mb v,\psi)$, one computes the
dressed propagator (the correlation function $G$) of the  "physical" fields,
\be
G({\mb x,y})= \langle\langle\psi({\mb x},t)\psi^\dagger({\mb y},t)\rangle\rangle=
C\int D\f D\mb vD\psi {\ } \psi({\mb x},t)\psi^\dagger({\mb y},t)\exp S(\f,\mb v,\psi).
\label{10}
\ee
in which the factor $C$ is defined by the normalization condition $ \langle\langle 1 \rangle\rangle=1.$

The correspondence between the original problem (\ref{eq}) and the field theory (\ref{9})
is not exact, since in the field propagator (\ref{10}) there are redundant diagrams containing closed loops
of the  $\psi \psi^\dagger $-lines. These diagrams were contributed to
 the correlation functions $\langle\f\f \rangle$  and  $\langle v_iv_j \rangle$  which we think of to be exact as defined in
 (\ref{korrfi}) and (\ref{4}).  In principle, these diagram  are to  be eliminated; one can simply say that
 the original problem is equivalent to the field theory (\ref{9}) except the diagrams containing
 the closed loops of lines of "physical" fields $\psi \psi^\dagger.$  This elimination does not
 affect the theory, since, first, we are not going to perform any functional variables transformations,
 which could mix up the redundant diagrams with those we need, and second, the renormalization
 group perfectly well permits such an   elimination of a certain class of diagrams.

We also note that all closed loops of  $\psi \psi^\dagger$-lines with no time derivatives in a cycle are
automatically equal to zero in  the theory (\ref{9}), since it contains     retarded and advanced  functions.
In particular,  from the  fact of elimination of all  diagrams containing  closed loops of oriented lines
it follows that

(i) correlation functions of any number  $n>2$ of the stochastic field $\f$ are trivial,
$$\langle\f(\mb x_1)\ldots\f(\mb x_n) \rangle=0;$$

(ii) correlation functions of any number  $n>2$ of the stochastic field $\mb v (\mb x, t)$ are trivial,
$$\langle\mb v (\mb x_1, t)\ldots\mb v (\mb x_n, t) \rangle=0;$$

(iii) any  mixed correlation functions of  stochastic fields  $\f(\mb x)$ and $\mb v (\mb x, t)$ are trivial,
$$\langle\f(\mb x_1)\ldots\mb v (\mb x_n, t) \rangle=0.$$

Furthermore, in the model (\ref{9}), the odd multipoint correlation
functions of  scalar field vanish, while the even single-time
functions satisfy  linear partial differential equations. The solutions for the
pair correlation function  can be  obtained explicitly in analogy with  that of
 in the Kraichnan model of turbulence \cite{11}.

Finally,  it is essential that the action (\ref{9}) is invariant under the following field
transformations:
\be
\begin{array}{c}
\psi({\mb x},t) \to \psi({\mb x+\mb s},t+\tau), \f_b({\mb x},t) \to \f({\mb x+\mb s},t+\tau)-b(  x ),\\
{\mb v}_a({\mb x},t) \to {\mb v}({\mb x+\mb s},t+\tau)-{\mb a}(t),
\end{array}
\label{12}
\ee
with two parameters: an arbitrary vector function of time ${\mb a}(t)$ decreasing
at $t\to -\infty$,  ${\mb  s}(t)= \int^t_ {-\infty}  \mb a (t')dt' $ and an arbitrary scalar function $b(  \mb x )$
fading out at $| \mb x |\to \infty,$ $\tau(\mb x)= \int_{\xi > |{ \mb x-\mb x'}|\geq  w} b( \mb  x') d\mb x'.$

The integral turbulent scale  as well as the dissipative length   are taken into account in the
model (\ref{9}) by the parameters of  infrared cut off    $m$   and ultraviolet
cut off   $\Lambda.$ The Ward identities which express the invariance of the model (\ref{9})
under the transformations (\ref{12}) guarantee that all UV singularities of diagrams
are subtracted out in each order of perturbation theory, so that all correlation functions
have  finite limits in $l_d\to 0$ and do not depend on $l_d$ in the inertial range.
In the region $m\to 0$ there are strong  infrared singularities (powers of $m$)
in the perturbation theory. However, these singularities are not related  to  the dynamic
interactions of eddies which form the spectra of the propagator $G$ but to the kinematic effect
of dragging of small eddies by the large ones \cite{a}. This effect can be
eliminated completely by   taking of the  frame of reference which moves with an  arbitrary
speed of the large-scale eddies.

Since these  singularities do not contribute to the spectrum for $G$, we will not
take them into consideration, assuming that the frame of reference  moves with some arbitrary
speed $v$  along the direction of flow. By the way, the solutions for concentration
is to be a traveling wave with the time argument $t-r\varepsilon/v,$ as it was  proven in \cite{4}.

\subsection{Dimensional Counting.  Action Functional with a  "Time-diffusive" Term
Included}

 We start now with  the dimensional analysis of   model (\ref{9}).
 Since the model (\ref{9}) is of two scales, one can introduce two independent canonical
dimensions to each quantity $F$ in the theory  (the momentum dimension, $d^k_F,$ and
the frequency dimension, $d^\omega_F$). Then, based on  $d^k_F$ and $d^\omega_F$,
on can introduce a total canonical dimension $d_F=d^k_F+2d^\omega_F$ (in the free theory,
$\partial_t\sim\Delta$). Assuming $d^k_k=d^\omega_\omega=1$, one finds out the  dimensions
of quantities in (\ref{9}) (see Tab. 1). Nonrenormalized parameters  in the Tab. 1 are supplied by the index $"0"$.

 Superficial {\em ultraviolet} (UV) divergences,  whose removal requires counterterms,  can present only in those
 Green's functions $\Gamma$, which canonical dimension
 \be
 \delta_\Gamma=d+2-d_\phi N_\phi
 \label{irasx}
 \ee
 is a nonnegative number, \cite{Col}. Here, $N_\phi$ is the number of corresponding fields entering
 the function $\Gamma $ and $\phi =\{\f, \mb v, \psi\}.$

 In the model (\ref{9})  the derivative $\partial$ at the vertex
  $\psi (\mb v\partial)\psi^\dagger$ can be moved onto the   field $\psi$
by virtue of  the vector field $\mb v$ is transversal. Similarly, the time derivative
$\partial_t$  at the vertex   $ c^{-1} \psi \f \partial_t \psi^\dagger$  can be moved onto
the field $\psi$ by virtue of the field $\f$ does not depend on time.
It decreases effectively  the real index of divergence: $\delta^{eff}_\Gamma =\delta_\Gamma
-N_{\psi}$.

From the dimensions  in the Tab. 1, taking into account the auxiliary considerations
(i)-(iii) of the previous Section as well as those of discussed above, one can find out that
for any $d$ superficial divergences can only exist in the 1-irreducible function
$\langle \psi\psi^\dagger\rangle$ for which $\delta_\Gamma=2, $ $\delta^{eff}_\Gamma=0.$
The corresponding counterterms must contain either two symbols $\partial$ and is therefore
reduced to $\psi\Delta\psi^\dagger$ or two time derivatives along with the squared $c^{-1},$
  $ \nu c^{-2} \psi\partial^2_t \psi^\dagger. $ Inclusion of these counterterms into the action functional
  (\ref{9}) cannot be reproduced by a simple multiplicative renormalization because of
the latter counterterm is not formally present in the action functional (\ref{9}).

To ensure the multiplicative renormalizability, we introduce  an additional term
$\nu_0c^{-2}_0\times $  $ \psi\partial^2_t \psi^\dagger$ into  (\ref{9})
supposing  $c_0$ to be a bare parameter; it leads to a renormalized action functional
of the form:
\be
\begin{array}{c}
S_R(\f,\mb v,\psi)= \int d{\mb x}dt\left[-\frac 12 \f D^{-1}_{\f} \f
-\frac 12 \mb vD^{-1}_v\mb v+
\psi^\dagger\left( \partial_t\psi + Z_1\nu\Delta\psi+Z_2 \nu c^{-2} \partial^2_t\psi + \right.\right.\\
+\left.\left.
 c^{-1} \f\partial_t\psi + (v\partial)\psi \right) \right] + C.C.
\end{array}
\label{sr}
\ee
The inclusion of the new term proportional to $c^{-2}$
 corresponds to  adding a term of the type $\nu c^{-2}\partial^2_tu(\mb x,t)$ into (\ref{eq}).
Notice that the inclusion of the  $\nu c^{-2}\partial^2_tu(\mb x,t)$-term
(a "time-diffusive" term) into the diffusion equation
in the presence of  axial depending porosity for the first time was proven in \cite{4}
for the case of  particular stochastic processes originating  the large scale porosity fluctuations.
 Our approach allows to justify   this result of \cite{4} and to
 generalize it  for any stochastic  law for a porous structure.
The coefficient  $\nu  c^{-2} $ is analogous to
the   time parameter $ \tau$ which was introduced in \cite{4} to
characterize time of the diffusion process measured in the time scale of the
carrier  fluid.

 Taking into account the new term in the diffusion equation (\ref{eq}),
 one obtains instead of (\ref{7}) in the limit $ (k\xi)^2\gg 1$
 \be
 \langle \psi({\mb k},t)\psi^\dagger({ - \mb k},t')\rangle_0 \simeq
   c{\ } e^{ -ck|t'-t|}/2\nu k.
 \label{p}
 \ee
 The damping of correlations in (\ref{p}) as $c\to 0 $ ($\xi \to \infty$)
 is an essential feature of the discussed approximation.

However, we may investigate the initial  theory with  no
such an addition, if we solve the RG equations obtained in extended model
with the initial condition $c^{-2}_0=0,$ {\em i.e., } if we suppose that  either
the diffusion process goes up extremely fast or the
correlation length $\xi$ is very short  due to  high dense packed bed.
 In both cases the domain of discussed microscopic phenomena
happens to be  small enough and the fluid flow behavior
in the packed bed column may be well investigated within
the framework of usual porous medium theory dwelling on the
constant averaged porosity.

In presence of the time-diffusive term
 the solutions (\ref{6}) exhibit a decay in time which rate is determined by the
singularity of  propagator $G,$ $k^2=\chi^2,$ where $\chi^2=\nu\w^2/c^2$
in Fourier representation is analogous to the  squared relative refractive
index in optics.  In a simple pole approximation for $G^{-1}=k^2-\chi^2$ the Bessel
function in (\ref{6}) can be replaced by its asymptotics at large $kr$
that yields the large distance asymptotics for $u(r)$:
$$  u(r)=\frac  i{2\chi}(\frac {\chi}{2\pi i r})^{(d-1)/2}.$$
In this case $Im \chi$ defines the extinction coefficient
$Im \chi\sim 1/l_0,$ where $l_0$ is the damping length.

\subsection{ Renormalization-Group  Equations,
Critical Scaling, and Empirical Laws of Kolmogorov and Richardson}

The  UV-divergences (in our case poles in $\epsilon$ and $\delta$  in
diagrams) of the extended model considered (with a "time-diffusive" term
included) are removed by the multiplicative
renormalization procedure. It amounts to the following: the initial  action $S(\Phi)$
is referred to as nonrenormalized, its parameters and coupling constants are referenced to as bare; these
 are considered as some functions  (remaining to be determined) of new renormalized
 parameters and coupling constants.

The   renormalized action functional (\ref{sr}) is a function of renormalized
coupling constants and parameters:
\be
\begin{array}{c}
c_0=cZ_c, \quad g_0=g M^{\epsilon}Z_g, \\
\eta_0=\eta M^{2\delta}Z_{\eta},\quad  \nu_0=\nu Z_{\nu},
\end{array}
\label{ren}
\ee
where  all renormalization constants $Z_{a}$ are the functions of two
independent   quantities $Z_1$ and $Z_2$:
\be
\begin{array}{c}
Z_{\nu}=Z_1, \quad Z_g= Z^{-1}_1, \\
Z_{\eta}=Z_1^{-5}Z_2^2, \quad Z_c=Z_1^{1/2}Z_2^{-1/2},
\end{array}
\label{z}
\ee
which can be calculated within the framework of diagram technique.
 We choose the simplest form of  subtraction scheme where the
divergences are presented as the bare poles in regularization parameters $\epsilon $
and $\delta$ (so called minimal subtraction scheme (MS));
$M$ is the renormalization mass parameter,
$g$, $\eta,$ and $\nu$ are renormalized analogues of the bare parameters $g_0$, $\eta_0,$ and $\nu_0.$
$Z_a=Z_a(g,\eta, c, \epsilon,\delta, d)$ are the renormalization constants.
The relations (\ref{ren}) and  (\ref{z}) result  from the absence of renormalization
for the nonlocal contributions  of  $D _{\f}$ and $D _v$ in the action functional
(\ref{sr}), so that $g_0\nu_0=g M^{\epsilon}\nu$ and   $\eta_0\nu_0^3/c_0^{-2}=M^{2\delta}\eta\nu^3/c^{-2}.$
No renormalization of the fields $\psi$ and $\mb v$  and  "masses"  are required, {\em i.e. }
$Z_{\mb v}=Z_{\psi}=1$  and $M_0=M,$ $m_0=m,$ $Z_m=Z_M=1.$

The only field $\f$   requires renormalization, $\f=\f_RZ_{\f},$ $Z_{\f}=Z_1^{1/2}Z_2^{-1/2},$
so that the renormalized Green's function $W^R$ meet the relation
\be
W^R(g,\eta, c, \nu, M)Z_{\f}^{N_{\f}}=W(g_0,\eta_0, c_0, \nu_0).
\label{cf}
\ee
$W^R$ are UV-finite function (  they are finite in the limits $\epsilon\to 0,$ $\delta\to 0$)
for fixed parameters $a.$

The RG equations  are written for the functions $W^R$ which differ from the initial
$W$ only by normalization and, thus, can be used equally validly for   critical scaling analysis.
To derive these equation one can note that  the requirement of eliminating singularities does not
determine the functions $e_0=e_0(e,\epsilon,\delta),$ $e=\{g, \eta, c, \nu \},$  uniquely because of the value of $M$ is not fixed
by any physical condition. Variation of $M$ for fixed values of bare parameters $e_0$ leads to variations of $e$ and
$Z_e,$ $Z_\f.$ Following the standard notation, we denote by $D_M$ the differential operator
$M\partial_M$ for fixed $e_0.$ Applying it on both sides of (\ref{cf}) leads to the basic
RG equation
\be
\left[D_M+\beta_g\partial_g+\beta_c\partial_c+\beta_{\eta}\partial_{\eta}-\gamma_{\nu}D_{\nu}+n_{\f}\gamma_{\f}\right]W^R=0,
\label{rg}
\ee
where we have used $D_x\equiv x\partial_x$ for any parameters of the renormalization theory; for any $Z_i$
\be
\gamma_i\equiv D_M\ln Z_i, \quad \beta_{\alpha}\equiv D_M\alpha, \quad \alpha \equiv \{g, \eta, c\}.
\label{rgf}
\ee
These identities determine the $\beta-$functions of the theory considered,
\be
\begin{array}{l}
\beta_g=-g\left(\epsilon-\gamma_1\right), \\
\beta_{\eta}=-\eta\left(2\delta-5\gamma_1+2\gamma_2\right), \\
\beta_c=-\frac c2\left(\gamma_1-\gamma_2\right).
\end{array}
\label{beta}
\ee
and the anomalous dimensionalities $\gamma_i.$
One calculates the renormalization constants $Z_1$ and
$Z_2$ from the diagrams of perturbation theory (for example, up to one-loop
order from the diagrams depicted in (\ref{8}))  and  then $\gamma_i$ , $\beta_{\alpha}$
functions.  All  $\gamma$- and $\beta-$
functions are constructed as   series in $g$, $\eta,$ $c,$ and  the functions $\gamma_i$
depend on neither $\epsilon$ nor $\delta.$

Actually,  the only reason for calculation of renormalization constants
in the theory considered is to prove   the existence of  {\em infrared} (IR)-attractive fixed
points $\{g_*,\eta_*,c_*\}$ of the RG equations (\ref{rg}) in the space
of renormalized   charges  such that
$\beta_{\alpha}(g_*,\eta_*,c_*)=0$ and $\w_{\alpha\zeta}\equiv \partial_{\alpha}\beta_{\zeta}(g_*,\eta_*,c_*)>0.$
The values of $\gamma_i(g_*,\eta_*,c_*)$ then can be found
from (\ref{beta}) exactly apart from diagram calculations, which are, in fact, pretty standard.

There is a line of  IR-fixed points  in the model discussed:
\be
g_*={\cal A} \epsilon,\quad \eta_*={\cal A}\delta,\quad   \forall  c_*, \quad {\cal A} = (4\pi)^{d/2} \Gamma(d/2)d (d-1)^{-1},
\label{fp}
\ee
  which
exists in the physical region $\{g,\eta,  c\}>0.$
Along this line
\be
{\gamma_1}_*=\epsilon,\quad {\gamma_2}_*=2\delta/3,
\label{fpg}
\ee
with no corrections of order $\epsilon^2,$ $\delta^2$ and so on.

There are no new critical exponents in the model considered
since the critical dimensionalities of $\f$, $c$, and ${\mb v}$ are
just equal to the Kolmogorov's value $-1/3$ as it defined from
the phenomenological assumptions (\ref{hyp}) and (\ref{4}), and
the dimensionality of the field $\psi$ is equal to its canonical dimension
 since it has no anomalous corrections ($\gamma_{\psi}=0$),
 \be
\Delta_{\f}=\Delta_c=\Delta_v=-1/3,\quad  \Delta_{\psi}=d/2,\quad \Delta_{\{\psi^{\dagger}\psi\}}=d.
\label{ind}
\ee
The results  expressed in the latter relation (\ref{ind}) can be applied
to the study of relative diffusion, {\em i.e.,} spreading of an admixture cloud consisting of a large number of particles,
\cite{Monin}. If we could label one of the particles in the cloud at time $t=0$ in the source of a frame $\mb x=0$,
then the effective radius of the cloud $R$ in the moment $t>0$ satisfies the relation
\be
R^2=\int d\mb x {\ }\mb x^2 \langle\psi^{\dagger}(\mb x, t)\psi (\mb 0, 0)\rangle.
\label{Rich}
\ee
Taking into account that $\Delta_R\equiv -1$, from (\ref{ind}) and (\ref{Rich})
one can obtain that $\Delta\left[dR^2/dt\right]=-2-\Delta_t,$ where
$\Delta_t$ is the Kolmogorov's dimensionality of time $\Delta_t=-2+\gamma_1=-2/3$.
It leads to $dR^2/dt\sim R^{4/3}, $ which was to be shown in the Richardson's
Four-Thirds Law, \cite{Monin}.

From the first relation  in (\ref{ind}) one can easily derive the well-known Kolmogorov's Law
of Five-Thirds, \cite{Monin}.

We should stress that the  indices (\ref{ind}) as well as their consequences
(Richardson's and Kolmogorov's Laws) are the universal features
of the model (\ref{sr}), {\em i.e.,} they  do not depend on the certain positions of
fixed points of the RG-transformation of variables in the domain of IR-stability.
Physically, this fact means that these properties do not depend on the parameters of the model
$\{g,\eta, c, \nu\}.$ However, the scaling function ${\cal F}$ for the propagator $G$
(and consequently the amplitude factors in the Kolmogorov's and Richardson's Laws,
{\em i.e.,} the Kolmogorov constant and the effective diffusivity coefficient)
do depend on $\{g,\eta, c, \nu\}.$ We discuss it in more details later on.

Another important note should be made on the role of   parameter $c $ for  the static
({\em i.e., } of equal time) spectra. As we have mentioned  before the parameter
$c$ determines the diffusion rate in the  microscopic scale by    means of
porosity  fluctuations.  However, the results (\ref{fp}) and (\ref{ind})
demonstrate that $c$ does not affect the universal properties of the static quantities of   model
but $c$ can only change the  amplitude factors of empirical laws.

 \subsection{Solutions of RG-Equations, Asymptotics for the correlation function $G$
 and the Extinction Coefficient $\chi$}

Now we derive  the solutions of RG differential equations (\ref{rg}) which
give the IR - asymptotics ($r\gg w $) for $G$.
 In renormalized variables the propagator $G$ can be expressed in the form
\be
G^{-1} =-k^d R(s,g,\eta,c, y, z ),
\label{g}
\ee
where $R$ is some function of dimensionless arguments
\be
s\equiv k/M,\quad  y\equiv \w/\nu M^2,\quad z\equiv  \chi^2/ M^2,
\label{arg}
\ee
which meets the RG-equation
\be
\left[ -D_s+\beta_g\partial_g+\beta_{\eta}\partial_{\eta}+
\beta_c\partial_c -(2-\gamma_1)D_y+(2\gamma_1-\gamma_2)D_z\right]R=0.
\label{rgr}
\ee
The equation (\ref{rgr}) can be solved and the function $R$ can be found out.

The general solution of (\ref{rgr}) is an arbitrary function of the first integrals
$\bar e =\{\bar g,\bar \eta,\bar c,$ $\bar y,\bar z\}$ which number is one less then
the number of arguments of $R$ in (\ref{g}). They can be founded from the
system of equations
\be
\frac{ds}{s}=\frac{d\bar g}{\beta_g(\bar g)}=\frac{d\bar \eta}{\beta_{\eta}(\bar g,\bar \eta)}=
\frac{d\bar c}{\beta_c(\bar g,\bar \eta)} =-\frac{d\bar y}{\bar y(2-\gamma_1)}=\frac{d\bar z}{\bar z(2\gamma_1-\gamma_2)},
 \label{fi}
\ee
supplied by some normalization conditions for $\bar e$. We use the  standard ones,
 \be
\bar e_i (s=1,e_i)=e_i.
\label{norm}
\ee
Traditionally, the  first integrals associated with coupling constants are
called  invariant or running   charges.  In spite of the model (\ref{sr})
is of three charges the relations between them are so simple that one
can (at least up to one-loop order approximation) solve (\ref{fi})
\be
\ln s=\int^{\bar g}_g\frac {dx}{\beta_g(x)}=\int^{\bar \eta}_{\eta}\frac {dx}{\beta_\eta(x)}.
\label{sfi}
\ee
These integrations can be brought about explicitly.
For invariant charges it, in particular, leads to  the   expressions
\be
\bar g(s,g)=\frac{\epsilon g}{\epsilon s^{\epsilon}+{\cal A}g(1-s^{\epsilon})},\quad
\bar \eta(s,\eta)=\frac{2\delta \eta}{2\delta s^{2\delta}+3{\cal A}\eta(1-s^{2\delta})}.
\label{ic}
\ee
Taking into account the normalization condition (\ref{norm}) and
the relations (\ref{sfi}), for any solution $R(s,g,\eta,c, y, z )$ we have
\be
R(s,g,\eta,c, y, z )=\left(\frac{s^{\epsilon}\bar g}{g}\right)^{-2/3}\left(\frac{s^{2\delta}\bar \eta}
{\eta}\right)^{-2/3}\left(\frac {\bar c}{c}\right)^{-2}
R(1,\bar g,\bar \eta,\bar c,\bar  y,\bar  z ).
\label{eqs}
\ee
Note, that the scaling function $R(1,\bar g,\bar \eta,\bar c,\bar  y,\bar  z )$ is not
fixed by   RG-equations and is usually calculated   in the framework of  diagram
technique.

 Under the renormalization group transformations  the canonical degrees of freedom are
 replaced by the scaling ones including anomalies. For the
 scaling  asymptotics $s\to 0$ of (\ref{eqs}) it  is expressed in the  fixation
 of the values of invariant charges   on their  values in the fixed points
 of RG-transformation (\ref{fp}),
 \be
 \bar g(s,g)\to g_*,\quad  \bar \eta(s,g,\eta) \to \eta_* , \quad s\equiv k/M\to 0.
 \label{c}
 \ee
In the practical problem it would rather be convenient to express the asymptotics
of  correlation function $G$ by virtue of   bare  parameters $e_0,$
using the relations of renormalization constants (\ref{ren}) and  (\ref{z}). Finally, one has
 the asymptotics $s\to 0$ in the following form
 \be
G^{-1} \simeq  Ak^d,
 \label{as}
 \ee
where $A=-\left(\frac  {g_0}{g_*}\right)^{-2/3}\left(\frac{ \eta_0}
{\eta_0}\right)^{-2/3}\left(\frac {c_0}{c_*}\right)^{-2}
 R(1,\bar g,\bar \eta,\bar c,\bar  y,\bar  z ).$

The essential feature of the model (\ref{sr}) is traced in
the dividing out of renormalization constants in the
IR-asymptotics of Green's function as well as disappearing of
the renormalization mass parameter $M$, which functions
in (\ref{as}) are commended to the bare parameters   $g_0,\eta_0, c_0, \nu_0.$

For the extinction index $\chi$  from (\ref{fi}) we have
\be
\bar \chi^2=\chi^2\exp\int^{\bar g}_gdx\frac{2\gamma_1(x)}{\beta_g(x)}-\chi^2
\exp\int^{\bar \eta}_{\eta}dx'\frac{ \gamma_2(x')}{\beta_{\eta}(x')}.
\label{chi}
\ee
In the analogous manner one can obtain the relevant scaling asymptotics:
\be
  \chi^2\simeq _{s\to 0} k^{8/3}\left(\frac  {g_0}{g_*}\right)^{2}\left(\frac{ \eta_0}
{\eta_0}\right)^{-1}\left(\frac {c_0}{c_*}\right)^{-2} \bar \chi (1,\bar g,\bar \eta,\bar c ),
   \label{aschi}
\ee
which corresponds to the following damping time-spectrum for the dynamic solutions $G$,
\be
G(t-t')\sim \exp -a_0 k^{2/3}|t-t'|,
\label{ts}
\ee
where $a_0= \left(\frac  {g_0}{g_*}\right)^{2}\left(\frac{ \eta_0}
{\eta_0}\right)^{-1}\left(\frac {c_0}{c_*}\right)^{-2}.$
Using (\ref{6}), one can compute the asymptotic value of  static
concentration spectrum (with no time dependence)
in the case of strong axial porosity fluctuations in the three dimensional space
 \be
u(r)\sim \frac 1r.
\label{conc}
\ee
 This asymptotics replaces the gaussian one for the region much large
 then $w$, so that the ordinary gaussian spectrum  for this region is to  be replaced
 by  $k^{-2}$  for $r\gg w$, see Fig. 2.

\section{Renormalization and Critical Dimensions of
Composite Operators $c^{-n}\f^n(\mb x)$}

We have considered the model (\ref{sr}) which corresponds to
an unphysical version of the packed beds problem. Now we
are going to generalize the developed technique to the case of
nongaussian fluctuative component of the
porosity field $\tilde\varepsilon.$

 In the Section 2 we have explained that
all   amendments due to deviations from the gaussian distribution for
$\tilde\varepsilon$ can be, in principle, taken into account (at least
they can be well-estimated) as a series of power-like composite operators
$\langle c^{-n}\f^n(\mb x)\rangle$ in the framework of renormalization-group approach for
the model (\ref{sr}).   Now we calculate the precise critical indices for
each power-like operator.

The canonical dimensions of  operators $\langle c^{-n}\f^n(\mb x)\rangle$ are trivial,
$d_F=-n d_c+n d_{\f}=0, \quad \forall n.$ From the Tab. 1. and (\ref{irasx})
one can see that  the only diagrams  with insertions of power-like
operators which contain the superficial divergences
have an arbitrary number of external "tails" of the $c^{-1}\f$-type.
However, at least one of such an external "tail" is attached either to the vertex
$\psi^{\dagger}(\mb v\partial)\psi$ or $ \psi^{\dagger}(c^{-1}\f\partial_t)\psi,$ so that
at least one derivative $\partial$ or  $\partial_t$ appears as an extra factor in the diagram, and, consequently,
the real index of divergence is necessarily negative.

This means that  all operators discussed require no counterterms, {\em i.e.,}
they are UV-finite. The same result can be readily reproduced
by consideration of the Ward identities which express the invariance of
the model (\ref{sr}) with respect to the field transformations (\ref{12}).
Generally speaking, the model (\ref{sr}) is invariant under the transformations
(\ref{12}) but the operators $\langle c^{-n}\f^n(\mb x)\rangle$ are not,
so that  they cannot have counterterms and cannot be UV-divergent.
 Finally, for the critical indices of  power-like operators $\langle\f^n(\mb x)\rangle$
one obtains the simple recurrent relation
\be
\Delta[\f^n]=-n/3.
\label{rr}
\ee
 Note that this relation was not initially  clear,  and it is a specific feature
 of the model (\ref{sr}) and those of similar type, \cite{12}.
Analogously, one has $\Delta[ c^{-n}\f^n(\mb x)]=0, \quad \forall n.$

The latter result exhibits that no one term of power series
for the fluctuative component of porosity field $\tilde\varepsilon$
can be omitted from the consideration in the equation (\ref{eq})
 since   their contributions are  of  equal importance.
 Thus, we need to consider the theory with all terms of the power series included
 in the action functional.

\section{  On the Solutions for Arbitrary $\tilde\varepsilon$}

The situation similar to that of the problem considered has place in the problem of chemically active scalar
 admixture \cite{hnat} and passive scalar admixture convection \cite{ant}.
Each term of the power-like series happens to be significant in the region we are interested in,
so that we cannot limit our consideration by neither  the first term $\sim c^{-1}\f$ as we did in
the Sec. 3 nor any finite number of terms in the power series for a porosity field function (for example,  (\ref{poros})).

Generally speaking, it would  lead to the investigation of the problem of an infinite number of dimensionless
coupling constants $\eta_n$, $n\to \infty$, {\em i.e., } to the renormalized action functional (which is  analogous to  (\ref{sr}))
of the form
\be
\begin{array}{c}
S_R(\f,\mb v,\psi)= \int d{\mb x}dt\left[-\frac 12 \f D^{-1}_{\f} \f
-\frac 12 \mb vD^{-1}_v\mb v+
\psi^\dagger\left( \partial_t\psi + Z_1\nu\Delta\psi+Z_2 \nu c^{-2} \partial^2_t\psi + \right.\right.\\
+\left.\left.
 c^{-1} \f\partial_t\psi + \sum _{n=2}^{\infty}c^{-n }  \eta_n \f^n\partial_t\psi /  n!+ (v\partial)\psi \right) \right] + C.C.
\end{array}
\label{act2}
\ee
 However,  in contrast with the problems  mentioned  any  of these new  terms,
 $c^{-n }  \eta_n \f^n\partial_t\psi /  n!$, does not contain any UV-divergences
 in accordance with the general dimensional analysis of Sec. 3.

This means that all the renormalization analysis of the theory (\ref{sr})
is still also valid for (\ref{act2}). In fact, the coupling constants $\eta_n$
are not the charges in theory (\ref{act2}). Any of the critical indices of the
theory (\ref{sr}) does not change in (\ref{act2}) and no one new critical
asymptotics  appeared.

The basic results (\ref{as}),
(\ref{aschi}), (\ref{ts}),   (\ref{rr}) are still secure for
any  power  series $c^{-n}\f^n$ in (\ref{act2}) {\em i.e.,} they
are valid for any porosity function distributed in the microscopical range.

The only difference  between the  various models of porosity  for  the certain  critical asymptotics
has place in the amplitude factors of  Green's functions. They do depend from
the certain initial values of the physical parameters of the problem.
We believe that it is the cause of a large dispersion of
the various experimental measurements in the
problem of the flow through the irregular packed beds.

\section{Conclusion  }

The final conclusion is that the  RG method  applied to the problem of packed beds
  with the diffusion coefficient which can arbitrary depend  from
the concentration of  marker in fluid flow  demonstrates the  existence
 of a scaling regime in the microscopical range where   porosity
   is  subject  to  the strong fluctuations.
 The relevant critical indices meet the empirical laws
 of Kolmogorov and Richardson for any porosity values. However,
the nonuniversal quantities of the theory, {\em i.e., }
the  amplitudes of empirical laws  are depended  on
the certain initial values of the parameters as well as the
porosity fluctuations. The fluctuative porosity forms the damping time-spectrum for the
dynamic correlation functions.
Finally, the solution  of nonlinear  equation of diffusion for the concentration
field $u(\mb x,t)$ is to be a traveling wave with an amplitude damping in space and time.
This fact distinguishes the
theory considered from the case of ordinary nonlinear  diffusion as well as
the case of  turbulent mixing of passive advecture.

We justified the result of \cite{4} on the  inclusion of a "time-
diffusive" term in the equation (\ref{eq}) for accounting of porosity fluctuation
and justified it for an arbitrary stochastic process originating
the porosity fluctuations.

\section{Acknowledgment}

We  thank G. Erochenkova for fruitful discussions.

\newpage

\vspace{6cm}

\begin{tabular}{|c|c|c|c|c| c|c|c|} \hline
\multicolumn{8}{|c|}{\bf Table 1. {\ } Canonical Dimensions of Quantities of the Theory} \\ \hline
\bf  F &  $\mb v, c, \f $ & $\psi$&   $g_0$& $\eta_0$ &  $ g ,\eta$& $m,\mu, M$& $\nu,\nu_0 $\\ \hline
 $d^k_F$&$-1$&$d/2$& $\epsilon$&$ 2\delta$&$0 $&$1$&$-2$\\ \hline
$d^\omega_ F$ &$1$&$0$& $0$&$0$&$0$&$0 $&$1$\\ \hline
$d_F$&$1$&$d/2$& $\epsilon$&$2\delta$&$0  $&$1$&$0$\\ \hline
\end{tabular}

 \newpage

 \begin{center}
CAPTION FOR FIGURES
 \end{center}

 FIGURE 1. On the definition of porosity.

{\small
The porosity is directly related to the size of the pores relative to the matrix.
If $P$ be a point inside a porous medium on the column axis
 surrounded by a layer of volume $V(l_i)$, one can define a ratio,
 $\varepsilon_i=\frac {V_v(l_i)}{V(l_i)}$
 in which $V_v(l_i)$ is the volume of void space within volume $V(l_i)$.
 }

 FIGURE 2.  The deviation of  dimensionless concentration spectrum from  gaussian for $r\gg w$.

{\small
Strongly developed fluctuations of the porosity and velocity fields lead
to anoumalously small fading off of concentration  on the distances
$r\gg w$  larger than the size of a pore.
 }

\end{document}